\documentclass[submission,copyright,creativecommons]{eptcs}
\providecommand{\event}{GCM 2019} 
\usepackage{underscore}           
\usepackage{graphicx}
\usepackage{stmaryrd}
\usepackage{float}
\usepackage{multirow}
\usepackage{color}
\usepackage{amsmath}

\title{Analysis of Graph Transformation Systems:\\ Native vs Translation-based Techniques}
\author{Reiko Heckel
\institute{Department of Informatics}
\institute{University of Leicester, UK}
\email{rh122@le.ac.uk}
\and
Leen Lambers 
\institute{Hasso Plattner Institut}
\institute{University of Potsdam, Germany}
\email{leen.lambers@hpi.de}
\and
Maryam Ghaffari Saadat 
\institute{Department of Informatics}
\institute{University of Leicester, UK}
\email{mgs17@le.ac.uk}
}
\def\titlerunning{Analysis of Graph Transformation Systems}
\def\authorrunning{R. Heckel, L. Lambers, M. Ghaffari Saadat}
\begin{document}
\maketitle

\begin{abstract}
The paper summarises the contributions in a session at GCM 2019 presenting and discussing the use of native and translation-based solutions to common analysis problems for Graph Transformation Systems (GTSs). In addition to a comparison of native and translation-based techniques in this area, we explore design choices for the latter, s.a.\ choice of logic and encoding method, which have a considerable impact on the overall quality and complexity of the analysis. We substantiate our arguments by citing literature on application of theorem provers, model checkers, and SAT/SMT solver in GTSs, and conclude with a general discussion from a software engineering perspective, including comments from the workshop participants, and recommendations on how to investigate important design choices in the future.  
\end{abstract}

\section{Introduction}

Logic-based methods have come a long way over recent years. Improvements in the usability and scalability of tools have led to significant advances in the automation of hard computational problems in software engineering. Automated formal verification, design space exploration, among others, require scalable solutions to constraint satisfaction or optimisation problems.
 
Analysis techniques for graph transformation systems pose a variety of hard computational problems. This includes techniques such as the execution, simulation or unfolding of systems, reachability analysis and model checking, the analysis of critical pairs, the verification and enforcement of graph constraints as invariants and the verification of systems based on a calculus of weakest preconditions. 
Many of these problems also arise in other contexts where state- and rule-based models or programs are analysed. One might expect that techniques adopted in verification more widely are also applicable to graph transformation systems, despite the fact that they are not particularly designed for our domain. This suggests a \emph{translation-based approach} where (typically logic-based) specifications are extracted and analysed in the target domain.

On the other hand, techniques providing solutions to our analysis problems are often based on theoretical results that take into account the specific features of graph transformations, such as their inherent non-determinism and concurrency, the complex non-linear structure of graphs, the properties of particular approaches and formalisations, and restrictions  including context freeness or monotonicity. As a consequence, the majority of existing analysis techniques are  \emph{native} ones, providing bespoke analysis tools for graph transformation systems and grammars. 

This paper aggregates  and elaborates the presentations and discussion at the eponymous session of GCM 2019. The objective is two-fold, to discuss the pros and cons of native vs. translation-based approaches to the analysis of graph transformation systems and, for the latter, understand some of the design choices influencing their qualities such as selecting the logics and tools, choosing an encoding, etc.  In Sections~\ref{sec:problems} and~\ref{sec:analysis} we provide an overview of a range of analysis problems and solution techniques. These two sections are based on a classification of the state of the art in this area in a forthcoming book on Graph Transformation for Software Engineers co-authored by Reiko Heckel and Gabi Taentzer. 
Then we review and compare native vs. translation-based solutions to analysis problems from the literature  and discuss reported experiments aimed at evaluating these solutions. Subsequently, we consider, in particular, literature on the use of theorem provers, model checkers, SAT and SMT solvers, which have seen some of the most impressive recent advances in technology.
 We conclude with a discussion from a more general software engineering perspective, including comments from the audience, and a general recommendation on how to investigate important design choices further.

\section{Analysis of Graph Transformation Systems}
\label{sec:problems}

\subsection{Usage and Semantics of Graph Transformations}

Graph transformations can be used in a variety of contexts and for different purposes, which often raise  specific analysis questions and require dedicated solutions. We distinguish three usage categories. 

Firstly, graph transformations can be used to define or characterise a set of graphs, i.e., a \emph{graph language}. A set of graph transformation rules with a start graph then act as \emph{graph grammar}. The idea is to consider as members of the language all graphs that can be generated from the start graph by the rules, in analogy to the formal definition of textual languages by means of string grammars. 


Secondly, graph transformations can be used to transform input graphs into output graphs; in other words, to program a \emph{graph relation}. Typical are translations between graphical languages, but also computations on graphs. Relations can be composed to reflect the sequential composition of computations. Such programmed graph transformation systems using control structures over rules can also fall into the next category, depending on whether we are interested in the overall I/O relation or their detailed behavioural semantics.

Thirdly, graph transformations can be used to describe the \emph{detailed behaviour} of a system. Here, individual rule applications represent actions that change the state of the system and can be observed, e.g., through their rule names. 
The semantics of such specifications is captured by labelled transition systems or other action- and / or state-based behavioural models. For example, in a \emph{graph transition system}, states are graphs and transitions correspond to transformation steps labelled by rule names, possibly augmented with parameters indicating to which nodes and edges the rule has been applied. 

\subsection{Problems of Interest}

Different semantic interpretations raise different questions depending on the use of graph transformations for describing a graph language, a relation between sets of graphs or a graph transition system.\footnote{This list concerns properties of individual graph languages. It does not include properties such as the closure of a certain class of languages under operations, e.g. union, intersection, complement, etc. 
}
\paragraph{Graph Languages}

Assuming a graph grammar $GG$ and its generated language $L(GG)$, the following properties are relevant.
\begin{itemize}
\item \emph{Membership:} Does the graph language $L(GG)$ contain a given graph $G$?
\item \emph{Inclusion:} Does the graph language $L(GG)$ contain another language $L'$, e.g., described by another grammar or declaratively by a type graph with constraints, or is $L(GG)$ itself a subset of another given graph language? 
\item \emph{Instance generation:} Can we enumerate the graphs $G$ in $L(GG)$ or sample this set randomly?
\item \emph{Non-ambiguity:} Does every graph in $L(GG)$ have a unique derivation up to reordering of independent steps, i.e., for any two derivations from the start graph $G_0$ to the same graph $G$, are they equivalent up to reordering? 
\end{itemize}


The last question is relevant as a prerequisite to parsing, a process by which we search for a derivation of $G$ by applying the rules in $GG$ backwards to reduce $G$ to the start graph $G_0$, thus solving the membership problem. The generating rules of the grammar are turned into \emph{inverse rules} for this purpose. If $G$'s derivation is unique, this provides us with information about the syntactic structure of the graph. Efficient parsing also requires the reduction rules to be terminating and confluent (i.e., deterministic up to choices that do not affect the end result). Non-ambiguity is not itself a property that can be stated based on the language semantics of sets of derivable graphs. It is actually  a property of the transition system, but is included under language properties here because of its relevance to parsing.

Non-ambiguity means that each well-structured diagram can be obtained by a unique sequence of transformations, usually up to reordering of independent steps. This is relevant because the derivation of a well-structured activity diagram reveals its hierarchical block structure. The problem of instance generation is relevant for testing or performance evaluation, e.g., of model transformation or analysis tools, where sample diagrams represent individual test cases. A parser for well-structured activity diagrams would produce derivations representing their hierarchical construction. 

\paragraph{Graph Relations}

A typical example of a graph relation is a translation between two modelling languages, e.g., from activity diagrams as input to Petri nets as output graphs. Assuming a transformation system describing a relation between two sets of graphs, we can consider the following properties.
\begin{itemize}
\item \emph{Functional behaviour:} Does the relation describe a function, i.e., does it associate each input graph with at most one output graph?

\item \emph{Totality:} Does the relation associate  to every input graph at least one output graph?
\item \emph{Injectivity:} Does the relation always map different input graphs to different output graphs?
\item \emph{Surjectivity:} Does it map an input graph to every output graph?
\item \emph{Correctness:} Is the relation consistent with the semantic interpretation of the graphs, or do output graphs satisfy a specific property $Q$ assuming that input graphs satisfy a suitable property $P$?
\end{itemize}

For a translation between graphical languages, functional behaviour and totality ensure that the mapping is well-defined as a total function, while injectivity and surjectivity imply that it is one-to-one and reaches all graphs in the target domain. 


\paragraph{Graph Transition Systems}

In the \emph{transition system domain}  we can ask questions about (sequences of) transitions and their interrelations. 
%
%
\begin{itemize}
\item \emph{Reachability:} Can a given graph, rule, or transition be reached from the start graph of the LTS? Can they be reached repeatedly? 

\item \emph{Invariants:} Do all graphs reachable from the start graph satisfy certain constraints or, dually, can we reach graphs violating such constraints?
\item \emph{Deadlocks:} Are there terminal states, i.e., without outgoing transitions? 
\item \emph{Planning and optimization:} Can we find a (good or optimal under a certain objective function) path in the LTS from a given graph to a graph satisfying certain properties?
\item \emph{Temporal properties:} Does the system satisfy certain safety and liveness properties, e.g., expressed by temporal logical formulas over paths through the system?
\item \emph{Termination:} Does the system only allow finite derivations?
\item \emph{Confluence:} Can every divergent pair of transition sequences be joined? If not in general, does this property hold locally for pairs of single transitions? Can we swap the order of consecutive steps in a sequence?
\end{itemize}

\section{Techniques}
\label{sec:analysis}
A variety of techniques are available to ensure the properties discussed above. 
\begin{itemize}
\item \emph{CDA:} Conflict and Dependency Analyses include static analysis techniques
(at rule-level rather than involving state graphs and transitions) to determine the possibility of conflicts or dependencies between rules. Critical pair analysis has its origins in term rewriting and has been generalised to term graph rewriting in \cite{Plu94,Plu99} and typed attributed graph transformations in \cite{HKT02,LBOST18,LSTBH18}.  Dependency analysis is an analogous technique applied to consecutive rather than branching transformations.

\item \emph{TA:} Termination Analysis also originates in term rewriting and was adapted first to term graph rewriting \cite{Plu98} and to graph transformation in~\cite{Ass00,BHPT05,VVE+06,DBLP:conf/birthday/Plump18}. It includes a range of techniques to establish the absence of infinite transformation sequences. Termination is undecidable in general, so these techniques are given in the form of sufficient criteria which, if satisfied, guarantee termination. 

\item \emph{WP:} Given a rule and a graph constraint intended as invariant, derives the Weakest Precondition of this constraint, as an application condition for the rule or a general graph constraint~\cite{HW95,EEHP04,HP09,EHLOG10}. In addition, one can check if the weakest precondition constructed is redundant since it is already entailed by existing application conditions or invariants. 

\item \emph{SSA:} State space analysis, including state space generation and model checking~\cite{RSV04,GMRZZ12,SV03,ABJKT10} but also techniques based on unfolding~\cite{BCK01,KK08} given a graph transformation system and a start graph, generates (an extract of) the transition system and then analyses it for satisfaction of temporal properties. Due to the ability to generate counter examples, model checking can also be used to construct sequences of transformations satisfying certain conditions, for example to generate test cases. 

\item \emph{GP:} Graph Parsing~\cite{BTS00,RS97,Min97}, means to reverse the rules of a grammar in order to solve the membership problem by reducing all graphs in the language to the start graph of the grammar. In the process, a derivation is produced which represents the syntactic structure of the graph. 
\end{itemize}

In most cases these techniques only provide incomplete answers to the questions stated earlier, e.g., in the form of sufficient criteria as for termination. Sometimes, combinations of techniques are required to address a single question. For example, to check if a transformation system with relational semantics implements functional behaviour, we have to establish its confluence. 
Confluence can be verified based on the computation of critical pairs. If all critical pairs are confluent and the graph transformation system is terminating, its transformation relation is a function. 
Without claiming completeness Table~\ref{default} indicates which technique could be used to address which questions.

\begin{table}[htp]
\caption{Analysis techniques to address  analysis problems}
\begin{center}
\scalebox{.7}{
\begin{tabular}{|r||c|c|c|c|c|}
\hline
				&	Conflicts \& Dep. 	& 	Termination  	& 	Weakest  & 	State Space  	&	Graph 	\\
				&	Analysis  	& 	 Analysis  		& 	Preconditions & 	Analysis 	&			Parsing \\

\hline
\emph{Language} &&&&&\\
\hline
Membership 		&				&				&			&	X 		& 	X		\\
Inclusion 			&				&				&	X		&	X		& 	X		\\
Instance Generation &	X			&				&	X		&	X		& 			\\
Non-ambiguity 		&	X			&	X			&			&	X		& 			\\
\hline
\emph{Relation}&&&&&\\
\hline
Functional Behaviour &	X			&	X			&			&	 X		& 	X		\\
Totality 			 &				&	X			&			&	 X 		& 			\\
Injectivity 			 &	X			&	X			&			&	 X		& 	X		\\
Surjectivity 		 &				&	X			&			&	 X		& 			\\
Correctness 		&	X			&				&	X		&	 X		& 			\\
\hline
\emph{Transition System} &&&&&\\
\hline
Reachability 		&				&				&			&	 X		& 	X		\\
Invariants 			&				&				&	X		&	 X		& 			\\
Deadlocks 		&				&				&			&	 X		& 			\\
Planning, Optimization &				&				&			&	 X		& 			\\
Temporal Properties &				&				&	X		&	 X		& 			\\
Termination 		&				&	X			&			&	 X		& 			\\
Confluence 		&	X			&	X			&			&	 X		& 			\\
\hline
\end{tabular}}
\end{center}
\label{default}
\end{table}

\paragraph{Language Properties}
Under certain prerequisites (non-ambiguity, functional behaviour of the reversed grammar)  \emph{membership} can be solved by graph parsing in an effective way, i.e., without backtracking. SSA can provide a partial solution, e.g., by generating a set of reachable graphs and checking if a given graph is in that set. If the language is finite (and small enough) this can be a complete (if inefficient) solution.   
Language \emph{inclusion}, for sets of graphs $L, L'$ with $L \subseteq L'$, can be sampled (tested) by any solution to the respective membership problems. If $L = L(GG)$ is the language generated by a grammar $GG$ and $L' = L(C)$ is the set of graphs satisfying certain constraints $C$, we can use constraint verification to verify that all graphs generated by $GG$ satisfy $C$, i.e., $L(GG) \subseteq L(C)$. 
The \emph{inclusion} question has a constructive version known as a \emph{filter problem}: Given $GG$ and a logical specification of $L$, how to derive a grammar $GG_L$ such that $L(GG_L) = L(GG) \cap L$. 

\emph{Instance generation} can be supported by SSA, generating graphs reachable from the start graph and returning them as counter examples to properties representing the negation of policies to determine which instances should be returned. Other approaches to sampling  ``interesting'' graphs include CDA and constraint verification, which generate minimal graphs demonstrating conflicts or dependencies or violating constraints. They can be checked for membership using any of the techniques above.
%
\emph{Non-ambiguity} can be verified by critical pair analysis on the set of inverse rules: if there are no critical pairs, the grammar is deterministic up to independence of transformations. In this case, all derivations a parser can return are equivalent.  Model checking can test non-ambiguity 
by trying to establish two different paths to certain graphs. 


\paragraph{Relation Properties}

As discussed above, \emph{functional behaviour} can be analysed by a combination of critical pair analysis, reachability and termination. 
Termination guarantees that a transformation relation produces a not necessarily unique result for all input graphs.  
Conversely, if functional behaviour can be established for the system obtained by reversing all rules, this can be used to show \emph{injectivity} of the original relation, while totality of this function establishes the surjectivity of the original relation. 

We distinguish syntactic and semantic \emph{correctness}. In the second category, if the semantics are described by operational or semantic mapping rules, \emph{mixed} confluence based on critical pairs between semantic and transformation rules can be used to show correctness. For syntactic correctness, we are interested to show that all graphs from the input set are mapped to syntactically correct graphs in the output set. This is also part of demonstrating that the function or relation implemented is total, and it can be achieved in part by showing that rules preserve or establish certain graph constraints.


%

\paragraph{Transition System Properties}

Many properties of states, transitions and paths in transition systems can be phrased as SSA problems. \emph{Invariants} can be verified by checking or enforcing the preservation of graph constraints. 
If the state space is described by a grammar that can be used for parsing, it can solve the \emph{reachability} problem. 
Temporal properties expressing safety conditions, such as the absence of unintended sequences, can be ensured constructively by imposing control structures.

\emph{Confluence} can be established by critical pair analysis in combination with termination and reachability.
Termination and confluence admit both analytical and constructive solutions. Termination by construction can be achieved by control structures such as layered graph grammars. A constructive approach to confluence  could either reduce non-determinism by suitable control structures or by completions adding rules to join diverging transformations.


\medskip
In general, graph languages, relations and transition systems are infinite. Therefore, many questions about them are only semi-decidable. That means, in order to answer them we seek sufficient criteria, or algorithms that over- or under-approximate the relevant properties, such as in the case of critical pairs: Their non-existence demonstrates that two rules can never create conflicting transformations but if a critical pair exists, the corresponding conflict may not be reachable from a given start graph. Reachability itself, like the membership problem, is a semi-decidable property.

\section{Native vs Translation-based Techniques}\label{sec:comparison}

In this section, we explore two different approaches to analyse graph transformation systems: native versus translation-based.  We start with defining more precisely what we actually mean with both terms and we illustrate these definitions with some examples from the literature following one or the other approach (cf. \autoref{sec:definition}).  We derive from these definitions some distinguishing characteristics that can help in guiding the selection of one or the other approach (cf. \autoref{sec:conceptual-comparison}).  We complement this conceptual comparison with an overview of experimental comparisons of both approaches that we have encountered in the literature for different analysis problems (cf. \autoref{sec:experimental-comparison}). We conclude this section with an evaluation of the question: Is there any empirical evidence backing up the conceptual comparison and how significant is it? Finally we discuss some challenges or open questions that arise from this evaluation. 

\subsection{Definitions and Examples}\label{sec:definition}

A \emph{native approach} to solving a graph transformation (GT) analysis problem is an approach where this problem serves directly as an input to a GT-specific solver. A \emph{translation-based approach} to solving a graph transformation analysis problem is an approach where this problem is translated to a logic-based specification, also called \emph{target specification}, in some logic-based domain, also called \emph{target domain}, where this problem is then also analysed. The target domain usually does not focus on graphs in particular. A \emph{hybrid approach} uses a mixture of the native as well as the translation-based approach to solve a graph transformation analysis problem. 

Let us have a closer look at \emph{model checking} for graph transformation as an \emph{example graph transformation analysis problem} in order to illustrate the above definitions. The model checking problem for GT can be formulated as follows: Is a specific liveness or safety property fulfilled in the graph transition system generated by a given start graph and a given set of graph transformation rules? The \emph{input} to the analysis problem consists of a start graph and a set of graph transformation rules as well as a liveness or safety property to be checked.  The \emph{output} of the analysis problem consists of the answer yes/no, whereas in the latter case the answer comes with a counterexample. An example approach following the \emph{native approach} to solving this problem is GROOVE~\cite{GMRZZ12}. This tool allows for feeding it directly with the above-described problem input and delivers the above-described output. The computations underlying GROOVE for solving the analysis problem are GT-specific. There exist several example approaches described in the literature~\cite{IsenbergSW13,BoronatHM09,BaresiS06} following the \emph{translation-based approach} to solving this problem by translating the latter to a target specification in first-order logic, rewriting logic, or relational logic, respectively.  Appropriate solvers for these target domains such as the SMT solver Z3~\cite{z3}, Maude~\cite{maude2007}, or Alloy~\cite{Jackson06}, respectively, are subsequently used to find an answer to the original GT analysis problem.  

\subsection{Conceptual Comparison}\label{sec:conceptual-comparison}

We can derive the following characteristics from the above definition of the \emph{native approach}:

\begin{itemize}
\item[N1.] \emph{No problem translation} necessary avoiding additional effort as well as potential errors due to translation.
\item[N2.] \emph{Promoted understanding of specifics} of graph and graph transformation analysis.
\item[N3.] \emph{Graph-specific optimizations} usually built-in.
\item[N4.] Structured support for \emph{different variants of graphs and graph transformation} promoting reuse of the commonalities of  underlying native techniques.
\end{itemize} 

On the contrary, we can derive the following complementary characteristics from the above definition of the \emph{translation-based approach}:

\begin{itemize}
\item[T1.] \emph{Problem translation necessary}, 
which might be a source of errors or misunderstandings.~\footnote{This translation may be automated reducing, in general, the source of errors or misunderstandings considerably.}  
\item[T2.]  \emph{Understanding of target domain and related solver(s) necessary} in order to obtain correct and useful target specification.~\footnote{If no automated translation is available, then the user needs this understanding, otherwise merely the developer of this automated translation needs it.}  
\item[T3.] \emph{Graph-specific optimizations} usually not built-in.
\item[T4.] \emph{Reuse experience and tool support} from target domain.
\end{itemize} 

Depending on the use case and based on these characteristics it might make sense to opt for one or the other approach. In addition to these characteristics described conceptually, we study more experimental comparisons of both approaches for some example analysis problems in the following. In particular, we will thereby focus the practical implications of the characteristics described conceptually here.  

\subsection{Experimental Comparison}\label{sec:experimental-comparison}

We have found a few \emph{experimental comparisons} in the literature with respect to following a native versus translation-based approach for solving particular graph transformation analysis problems.  In particular, we describe the main findings of such an experimental comparison for model checking graph transformation systems~\cite{RSV04}, constraint verification applied to pre- and post-condition reasoning~\cite{Pennemann2009}, constraint verification applied to invariant checking~\cite{BeckerBGKS06}, and constraint verification in the sense of satisfiability solving and  automated reasoning~\cite{SemerathNV18,SchneiderLO17}. 

We start with a generic description of the analysis problem at hand as well as giving a few pointers to example approaches solving the analysis problem following the native or translation-based approach. Then we report more in detail on the above-mentioned experimental comparison found in the literature and describe their practical findings with respect to the conceptual characteristics of each of the approaches. 

\paragraph{Model checking} We have described the model checking problem for graph transformation already in \autoref{sec:definition}, where we have listed some pointers to representatives of the native vs translation-based approach to solving this problem. Now we report on the \emph{experimental comparison}~\cite{RSV04} of the native approach followed by GROOVE~\cite{GMRZZ12} and the translation-based approach followed by CheckVML~\cite{SV03}, exploiting off-the-shelf model checker tools like SPIN~\cite{Holzmann97}. On the one hand, it is reported that GROOVE is able to ''Simulate graph production rules directly and build the state space directly from the resultant graphs and derivations. This avoids the preprocessing phase, and makes additional abstraction techniques available to handle symmetries and dynamic allocation.'', referring to characteristics N1 (No translation) and N3 (Built-in graph-specific optimizations) described in \autoref{sec:conceptual-comparison} in particular. On the other hand, it is reported that CheckVML is able to ``Encode graphs into fixed state vectors and transformation rules into guarded commands that modify these state vectors appropriately to enjoy all the benefits of the years of experience incorporated in existing model checking tools.'', referring to characteristics T1 (Problem translation necessary), T2 (Understanding of target domain and related solver(s) necessary) and T4 (Reuse experience and tool support) in particular. The overall conclusion sounds as follows ``CheckVML outperforms GROOVE if the dynamic and/or symmetric nature of the problem under analysis is limited, while GROOVE shows its superiority for inherently dynamic and symmetric problems.'', referring to characteristics N3 (Built-in graph-specific optimizations), T3 (No built-in graph-specific optimiztations) and T4 (Reuse experience and tool support) in particular. 


\paragraph{Pre- and post-condition reasoning}
The related graph transformation problem can be formulated as follows: Given an input graph satisfying a particular pre-condition, does the output graph generated by the given graph program satisfy the post condition? 
The \emph{input} to this analysis problem consists of a pre- as well as post-condition (in the form of graph conditions) and a graph program. The \emph{output} consists of the answer yes, no, or unknown, since this problem is undecidable in general. 
Solving this problem is usually performed with some kind of interactive theorem proving, where the user needs to specify, for example, loop invariants. 
A first example \emph{native approach}~\cite{HP09} is based on Dijkstra's approach to program verification and adapted to graph programs, in particular. A second example native approach~\cite{PoskittP12} is based on a Hoare-style proof system for graph programs. 
Two example \emph{translation-based approaches}~\cite{Strecker18,BrenasES18} translate the problem to a target domain like Isabelle/HOL~\cite{NipkowPW02} or description logics~\cite{Baader03a}, respectively.  

We report in particular on an \emph{experimental comparison} of a native and translation-based approach to this problem as described in the PhD thesis of Karl-Heinz Pennemann~\cite{Pennemann2009}. The native approach is based on a native theorem prover ProCon and SAT solver SeekSat for graph conditions, whereas the translation-based approach resorts to off-the-shelf first-order logic theorem provers and satisfiability solvers such as e.g. VAMPIRE~\cite{RiazanovV02} and DARWIN~\cite{BaumgartnerFT06}, respectively. The author reports that 
''ProCon and SeekSat are structure-specific in a constructive way. In contrast, theorem prover and satisfiability solver for general first-order logic necessarily consider arbitrary structures and have to be restricted by a set of axioms to a target structure which adds to the complexity of the problem.'' This illustrates the characteristics N1 (No problem translation necessary) and T1 (Problem translation necessary) of each approach, respectively. 
Moreover, he reports on the characteristic N3 (Built-in graph-specific optimizations) of the native solvers in the following way: ''An algorithm on conditions can and should use the fact that conditions make quantifications and statements in bulks, that is, a quantifier may introduce a number of elements. In this sense, conditions may have a lower logical complexity when compared to their translations in first-order logic.'' Moreover, experiments on several case studies in the thesis have demonstrated that the native solvers outperform off-the-shelf solvers from the target domain when it comes to efficiency, which can be seen as an illustration of N3 as well as T3 (No built-in graph-specific optimizations). 
In particular, he reports also on characteristic T2 (necessary understanding of the target domain) as follows: 
``For formulas, it remains open if the values of variables are equal or distinct, unless it is explicitly stated. If the nodes and edges of a graph condition are not distinct by their labels, inequations have to be introduced during the translation.'' Finally note that the native approach illustrates nicely characteristic N4 (support for different variants of graphs and GT), since the underlying theories and tooling are based on the framework of weak adhesive high-level replacement categories~\cite{EhrigEPT06} supporting these different variants. 

\paragraph{Invariant checking}
We formulate the related graph transformation problem as follows: Does each rule application via a rule of a given set of GT rules on a graph satisfying a particular graph condition lead to a graph satisfying this condition again?  The \emph{input} to this problem is a graph condition together with a set of GT rules. The \emph{output} consists of the answer yes, no, or unknown, since in general this is an undecidable problem.  
An example \emph{native approach} to solving this problem is presented by Becker et al.~\cite{BeckerBGKS06}, whereas a translation-based approach is described by König et al.~\cite{KonigE10}.  The latter approach is based on an approximation by Petri nets.

In particular, Becker et al.~\cite{BeckerBGKS06} describe not only a native approach to the invariant checking problem for GT (called explicit algorithm in the following), but present in addition an \emph{experimental comparison} with a translation-based approach (called symbolic algorithm in the following). The symbolic algorithm resorts to the relational programming language RML as target domain with the related solver CrocoPat~\cite{BeyerNL04}.  The authors in particular report on efficiency issues, illustrating evidence for characteristic N3 (Built-in graph-specific optimizations) as well as T4 (Reuse experience and tool support) in the following way, respectively: ``For the explicit algorithm, the combinatoric complexity of the rule/invariant pair has the most significant impact on the computation time. This explains why the pair goDC2 and noDC is a particularly easy case for the explicit algorithm, in spite of the size of the pair, as the number of possible intersections is constrained by a large number of positive edges.''\footnote{Note that goDC2 is a rule and noDC an invariant.}
and ``Speed-up due to the symbolic encoding can be extremely high for certain hard cases with a high number of nodes and edges.''. 

\paragraph{SAT-solving and automated reasoning}
The \emph{SAT-solving problem} 
for graph conditions can be formulated as follows: Does there exist a graph satisfying the given graph condition? The \emph{automated reasoning problem} on the other hand can be considered as complementary and can be formulated as follows: Do all graphs satisfy the given graph condition? The \emph{input} to both problems is a graph condition and the \emph{output} consists of the answer yes, no, or unknown, since in general both problems are  undecidable. There exist a number of example \emph{native} approaches as well as \emph{translation-based} approaches to both problems.  For example, Pennemann~\cite{Pennemann08} presents a native theorem prover, whereas Schneider et al.~\cite{SchneiderLO17} and Semeráth et al.~\cite{SemerathNV18} present native SAT solvers for graph conditions. Example translation-based approaches~\cite{KuhlmannHG11, GonzalezBCC12, SemerathVV16} map the SAT solving problem to target domains such as relational logic~\cite{Jackson06} and constraint logic programming. 

We first report on an \emph{experimental comparison} to SAT solving~\cite{SemerathNV18} of a native approach and translation-based approach using Alloy~\cite{Jackson06} concentrating on scalability of the corresponding solutions. In particular, the authors write the following conclusions from their experimental comparison, illustrating the characteristics N3 (Built-in graph-specific optimizations) as well as T2 (Understanding of target domain and related solver(s) necessary): ''Our graph solver provides a strong platform for generating consistent graph models which are 1-2 orders of magnitude larger (with similar or higher quality) than derived by mapping based approaches using Alloy with an underlying SAT-solver. Such a difference in scalability can only partly be dedicated to our conceptually different approach which combines several advanced graph techniques to improve performance instead of fine-tuning a mapping. However, it likely indicates fundamental shortcomings of existing mapping based approaches. Based on in-depth profiling we suspect that representing each potential edge between a pair of nodes as a separate Boolean variable blows up the state space for sparse graphs with only linear number of edges.''.  

We conclude with describing another experimental comparison to SAT solving~\cite{SchneiderLO17}, again of a native approach implemented in the tool AutoGraph and translation-based approach using Alloy~\cite{Jackson06} focusing efficiency as well as conciseness of the generated solutions. In particular, the authors write ``AutoGraph is capable of obtaining minimal, symbolic models, which allow for a straightforward exploration of further models whereas Alloy generates models for a given scope not necessarily determining minimal models. Also AutoGraph allows for the refutation of a given formula, which is not directly given in Alloy where non-existence of models is also bound to scope sizes. Hence, we conclude that AutoGraph computes in this sense stronger results compared to Alloy.'' and ``We observed for our running example comparable runtimes. However, as stated before, AutoGraph already returns stronger results by computing not only a minimally representable model, but a symbolic model.''. This illustrates on the one hand the characteristic N2 (Promoted understanding of specifics of graph and GT analysis) of the native approach and on the other hand the characteristic N3 (Built-in graph-specific optimizations) for the native and T4 (Reuse experience and tool support) for the translation-based approach.  

\subsection{Evaluation}\label{sec:evaluation-comparison}

The experimental comparisons studied in the literature and reviewed in \autoref{sec:experimental-comparison} demonstrate that each of the characteristics for the native versus translation-based approach as identified in \autoref{sec:conceptual-comparison} indeed play a role in practice. Each of the experimental comparisons basically showed which practical implications some of the conceptual characteristics have that can then account for a significant difference between a native versus translation-based approach. Therefore we suggest that consciously investigating the practical implications of each of the characteristics for each new use case, might help in guiding the decision between a native or translation-based approach. 

Open questions, discussion topics and challenges arising from this evaluation are the following:
\begin{itemize}
\item Is the list of characteristics from the conceptual comparison in \autoref{sec:conceptual-comparison} complete enough to be able to guide the choice between a native or translation-based approach for each use case? If not, do we need more, or also more specific characteristics, e.g. parametrized by the type of analysis problem? Will future experiments contradict some of the characteristics such that they would need to be rethought?
\item Why are some analysis problems currently addressed prevalently by native (or translation-based) approaches? Are there problems for which a native (translation-based or hybrid approach) would be more appropriate?  
\item Which target domains have been used for translation-based approaches and why are they appropriate for the given graph transformation analysis problems?
\item What can we learn from native versus translation-based approaches and experimental comparisons in the past for core computations such as e.g. the subgraph isomorphism problem?
\end{itemize}

\section{Use of SAT and SMT Solvers}
Solvers for boolean satisfiability problems (SAT) are generic tools that solve constraints involving only boolean variables. A SAT solver takes as input a propositional formula $\phi$ in conjunctive normal form (CNF) and outputs True if $\phi$ is satisfiable and False otherwise. Solvers for satisfiability modulo theories (SMT) are extensions of SAT solvers equipped with background theories to handle non-boolean variables as well. Examples of theories used in GTSs are the theory of real numbers, the theory of integers, and the theories of various data structures such as lists, arrays, bit vectors and so on. An SMT solver takes as input a ground first-order logic formula $\psi$ and a background theory $T$ and outputs True if $\psi$ is satisfiable with respect to $T$ and False otherwise.\\ 

An SMT solver has two components that are meant to combine advantages of abstract and domain-specific approaches: 1) a SAT solver for efficient and abstract reasoning, and 2) a theory solver for domain specific reasoning the result of which is passed on to the SAT solver. For example, given an SMT solver equipped with a theory of integers, a 32-bit integer variable in an SMT instance can be translated to SAT format using 32 bit variables with appropriate weights, and word-level operations such as 'plus' in SMT can be translated into lower-level logic operations on the bits in SAT. The relation of SAT- and SMT- solvers is analogous to that of assembly language and high-level programming languages, in that, in both cases the former is much more efficient but the latter is more developer-friendly and thus easier to work with.\\

Both SAT- and SMT- solvers are in essence generic tools to solve systems of equations and since many problems can be formulated as a set of equations, SAT/SMT solvers are quite versatile target domains in the translation-based approach. In this section we review a few applications of SAT- and SMT- solvers in Graph Transformation Systems (GTSs) .

\paragraph*{Example: Graph Matching using a SAT solver \cite{Rudolf1998}}{Graph matching is a basic step required for rule application in GTSs and plays a key role in almost all analysis techniques. The graph matching problem is essentially a subgraph homomorphism problem (NP-complete, exponential in the size of the left-hand side of the rule). In order to translate a graph matching problem for a rule $r$ into a SAT formula, each node/edge in the domain of the match of $r$ is represented by a unique variable, and each element in the co-domain of the match is represented by a value. The homomorphism condition can be expressed in the form of a propositional formula in terms of the variables. If the formula is satisfiable, the assignment produced by the SAT solver corresponds to a graph morphism in the GTS. An example is shown below where a graph morphism with its domain and codomain are encoded into a SAT formula:\\

\begin{tabular}{c c}
\begin{minipage}{.4\textwidth}
\includegraphics[scale=.3]{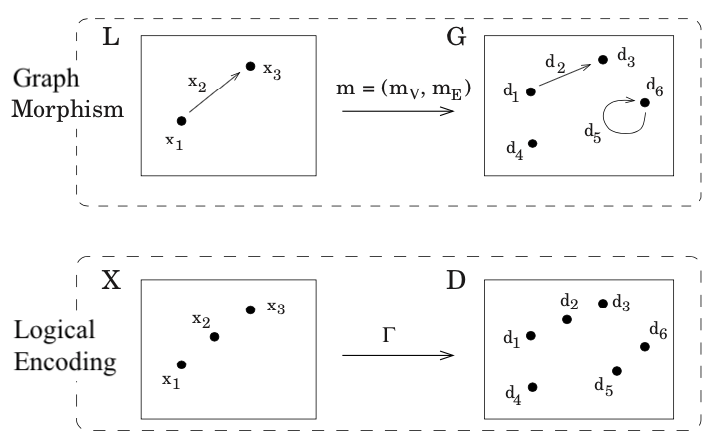}
\end{minipage}
&
\begin{minipage}{.55\textwidth}
{\footnotesize
\begin{align*}
node~variables & : & x_1, x_3 \\
edge~variable & : & x_2 \\
node~values & : & d_1, d_3, d_4, d_6 \\
edge~values& : & d_2, d_5  \\
edge~source~constraints & : &  (d_2, d_1) \in C^{src}_{(x_2,x_1)}\\
&  &  (d_5, d_6) \in C^{src}_{(x_2,x_1)}\\
edge~target~constraints & : &  (d_2, d_3) \in C^{tar}_{(x_2,x_3)}\\
 & &  (d_5, d_6) \in C^{tar}_{(x_2,x_3)}\\
solution~1 & : & \Gamma_1(x_1,x_2,x_3) = (d_1, d_2, d_3)\\
solution~2 & : & \Gamma_2(x_1,x_2,x_3) = (d_6, d_5, d_6)\\ 
\end{align*}}
\end{minipage}
\end{tabular}\\

In this example, morphism $m$ maps graph $L$ into $G$ and $\Gamma$ is the corresponding assignment in the logical domain from variables in $X$ to values in $D$. There are two nodes in $L$ represented by variables $x_1$ and $x_3$ in $X$ and the edge $x_2$ from $x_1$ to $x_3$ in $G$ is represented by variable $x_2$ in $X$. Every node/edge in $G$ represents a possible node/edge value in $D$. The homomorphism conditions make sure that for instance if $x_2$ is mapped to $d_2$, then its source $x_1$ is mapped to the source of $d_2$ which is $d_1$. In other words, the source and target mappings must be preserved by the morphism. Given the node and edge variables, their corresponding sets of possible values, and the homomorphism conditions, there are two SAT solutions in the form of instantiation of all variables with the given values such that the homomorphism constraints are satisfied: edge $x_2$ is either mapped to $d_2$ or $d_5$. In the former case $x_1$ and $x_3$ are assigned $d_1$ and $d_3$ respectively whereas in the latter $x_1$ and $x_3$ are both mapped to $d_6$.}

Translation to SAT/SMT can reduce average-case complexity of graph matching and thus that of the GTS analysis as a whole. More generally, the following are potential advantages of proper utilisation of SAT/SMT solvers in GTSs.
\begin{itemize}
\item \textbf{Efficiency:} instead of inventing new algorithms from scratch for graph matching or model checking, we can employ well-established SAT/SMT solvers which are highly efficient.
\item \textbf{Modularity:} due to independence of SAT/SMT algorithms from the concrete graph model, changing the model only requires adaptation of the translation step rather than reformulation of the matching or model checking algorithm. This is particularly helpful if an SMT solver is used with suitable background theories, e.g. a graph theory, to facilitate the translation step. 
\item \textbf{Generality, common across GTSs:} due to being abstract and general-purpose, SAT/SMT algorithms can be reused easily across various GTSs and enable a better comparison based on the parts that are specific to each GTS. This advantage differs from modularity above in that this concerns comparison of alternative algorithms whereas modularity is helpful in developing and changing a single algorithm. 
\end{itemize}

Analysis techniques for GTS to which SAT/SMT solvers have been applied include WP (Section \ref{sec:wp}) in a graph specification framework for verification (DrAGoM \cite{Dragom2019}), TA (Section \ref{sec:ta}) using Petri-net approximation and weighted type graphs (GREZ \cite{Grez2015}), and SSA tools (Section \ref{sec:mc}) such as Mini-SAT \cite{Kreowski2010}, a GROOVE extension  \cite{IsenbergSW13}, SGA \cite{Steenken2015, Steenken2011}), and instance generators Cartier \cite{sen2008combining}, Reflective Instantiator \cite{mcquillan2008metamodel}, and ASMIG \cite{Wu2013exploiting} for metamodels (Section \ref{sec:ig}).


\subsection{Strongest Postcondition}
\label{sec:wp}
DrAGoM (short for: Directed Abstract Graphs over Multiplicities) \cite{Dragom2019} is a prototype tool to handle and manipulate \textit{multiply annotated type graphs} originally defined in \cite{DBLP:conf/gg/Corradini0N17,Corradini2019specifying}. The main application of DrAGoM is to check invariants of GTS in the framework of abstract graph rewriting. 

Given a graph language, the tool constructs an abstract graph which specifies the strongest postcondition of the GTS and uses a materialization construction to extract a concrete instance of the left-hand side out of the abstract graph in every possible way \cite{corradini2019rewriting}. Since computing annotations for the rewritable materialization is a search problem of a high time-complexity, DrAGoM encodes the problem of annotation computation into an SMT formula by encoding nodes and edges as variables, declaring annotation functions with ranges of codomains specified by assertions, using the morphisms to compute sets of nodes and edges with specific requirements, adding constraints to enforce maximality of pairs of annotations, and encoding constraints for the elements not in the codomain of the morphisms. It then passes the resulting SMT formula to the external solver Z3. Whenever the solver finds a model, it produces a pair of annotations which yields a legal morphism and the pair is maximal in the sense that we exactly hit all desired bounds specified by the abstract rewriting step defined in \cite{Nolteanalysis}. The SMT formula is then extended to exclude all previously found models and checked for satisfiability until the formula gets unsatisfiable. In each iteration step another maximal pair is found. Afterwards DrAGoM performs a language inclusion check to verify if the strongest postcondition is already covered by the initial graph language.

In practice, DrAGoM uses an optimization where, in case of a universal quantification, the variable is substituted with every possible instance and the universally quantified formula is replaced by the conjunction of the resulting instantiated formulas. This is possible since we always quantify over finite sets.

\subsection{Termination Analysis}
\label{sec:ta}
The GREZ tool \cite{Grez2015} takes a GTS as input and tries to find a proof that it terminates or a proof that it does not. GREZ uses the double pushout approach on hypergraphs and runs a number of algorithms concurrently. It reports the result of the first of the algorithms that successfully finishes. Since the problem is undecidable in general, in many cases none of the algorithms will find a termination or non-termination proof.\\
GREZ employs SMT solvers with the theories of linear integer arithmetic and uninterpreted functions, using quantifier-free logic. The GREZ algorithms that require an SMT solver are:

\begin{itemize}
\item Petri-net approximation: Over-approximates the transition sequences of a GTS by a Petri-net. In other words, termination of the Petri-net implies termination of the GTS but not vice versa. Theorem 3 in \cite{VVE+06} specifies a linear inequality as the condition for non-termination of a Petri-net. An SMT solver is used to check whether this condition holds for the Petri-net approximating the GTS or not. 
\item Weighted type graph method: This technique uses type graphs to assign weights to graphs that strictly decrease in each graph transformation step. A weighted type graph is a type graph where each node and edge is given a weight in natural numbers. A GTS is proved to be terminating if there exists a weighted type graph which satisfies the required properties which can be encoded in an SMT formula in GREZ and passed on to an SMT solver along with the number of nodes $k$ to check whether a weighted type graph with $k$ nodes exists which satisfies the said properties. 
\end{itemize}

\subsection{Model Checking}
\label{sec:mc}
Model checking requires searching the possible derivations for counterexamples which is often computationally expensive due to nondeterminism of GTSs. A controlled search using SAT/SMT can considerably reduce the average-case complexity of model checking.\\

In \cite{Kreowski2010}, graph transformation units (encoding solution concepts) are translated into a propositional formula to feed into a SAT solver (Mini-SAT). The results confirmed that for established benchmark functions, target problems can be efficiently solved using the proposed approach. For all considered graphs the considered problem can be solved in very short run-time. In \cite{ermler2011graph}, the authors describe how a SAT solver can be integrated into the analysis of graph transformation units to facilitate the search for solutions to NP-hard problems. They have designed and implemented the prototypical tool SATaGraT (SAT solver assists graph transformation engine) and tested it on various problems including the job-shop problem. They found that their translation of graph transformation units into propositional formulas yields very large numbers of clauses although the formulas are of polynomial size. Hence, further optimizations are needed to obtain a SATaGraT behavior that is competitive with the benchmarks in the literature. 

In \cite{IsenbergSW13}, a bounded model checking method is proposed which is claimed to be the first attempt to encode the complete GTS verification problem directly to satisfiability checking. The aim is to use SMT solvers to unfold the transition relation up to a predefined boundary to find error states such as forbidden graph patterns or LTL formulas. The background theory used in the SMT solver is the theory of uninterpreted functions with equality. 

The authors provide a prototypical implementation as an extension to GROOVE which uses Z3. They observed that the overhead of using SMT instead of SAT is small for the used theory. Moreover, using an SMT solver increases performance by restricting the search space more quickly than a SAT solver. The order of encodings of transformation rules matters (w.r.t. solving time), particularly when having only one path containing some forbidden patterns in a large set of paths. Overall this technique can be useful for error finding in GTSs with infinite state spaces.\\

In \cite{Steenken2015, Steenken2011}, the authors address model checking of GTSs with infinite numbers of states via abstraction. A {shape} is a generalisation of graphs, encoded in three-valued logic in order to model uncertainty as \textit{potentially} true or false and label {summary nodes} which are equivalence classes of nodes sharing certain properties such as neighbourhood connections.

The authors of \cite{Steenken2015, Steenken2011} provide a prototypical implementation in Java composed of two parts, the Shape Graph Analyzer (SGA) and its accompanied SMTLib-library called SMTool, interfacing with solvers SMTInterpol and Z3. They use a Lazy State Space Construction (LSSC) to represent infinite state sets with finite shapes and transform the shapes in a sound abstraction of SPO semantics. According to their evaluations, modern SMT solvers are capable of solving large and complex models in a matter of seconds or minutes at most. Furthermore they found that, using their proposed methods, a vast majority of the shapes can be checked for feasibility in just a few seconds.\\

\subsection{Instance Generation} 
\label{sec:ig}
Generating instances for an abstract model, a.k.a. a metamodel, is generally defined as producing concrete, fully instantiated models that adhere to the abstract model both syntactically and semantically. Several approaches to instance generation, including \cite{semerath2019viatra, SemerathVV16, SchneiderLO17}, have been discussed in Section \ref{sec:experimental-comparison}. In what follows, we consider a few translation-based approaches to instance generation which use existing SAT/SMT solvers.\\

In \cite{Wu2013exploiting}, an approach is presented to generate instances of metamodels using an SMT-solver. They represent a metamodel by a bounded attributed type graph with inheritance ($ATGI_b$) based on which a finite universe of bounded attributed graphs $AG_u$ typed over $ATGI_b$ is constructed representing a superset of its instances. Then the nodes and edges in $AG_u$ are translated into quantifier-free SMT formulas and fed into an SMT solver. If the SMT solver outputs a successful assignment of values to nodes and edges, an instance of the metamodel is found. Otherwise, the metamodel is inconsistent in the current bounds and an instance may be found within larger bounds.

In \cite{wu2012metamodel}, the authors present a systematic literature review of instance generation techniques for metamodels. Their review includes citations to SAT/SMT-based approaches most of which interact with Alloy \cite{Jackson06}, a tool that uses bounded relational logic to describe a model and interacts with an external SAT-solver to find instances of and counter examples for a model in a finite search scope. A couple of such references that use SAT-solvers are \cite{sen2008combining, mcquillan2008metamodel} which present tools called Cartier and Reflective Instantiator respectively. A notable reference in their review is \cite{el2011relational} which utilises the ability of SMT-solvers to perform unbounded integer arithmetic operations by axiomatising Alloy specifications into SMT-instances (using a first-order logic) to prove Alloy assertions. \\

\subsection{Concluding Remarks}
To our knowledge, there are only a small number of graph transformation tools that use SAT/SMT solvers. However the studies cited in this section show that these tools are powerful in finding matches for rules in large graphs and navigating complex state spaces. SMT solvers in particular have been used in the analysis of GTSs due to utilising domain-specific reasoning alongside abstract SAT-solving.

\section{Discussion and Outlook}
In this section we reflect on the discussion at the workshop and attempt to synthesize some of the findings of the previous sections. Where comments refer to remarks made at the GCM session, we refer to participants by name.
%
%
Among the analysis approaches considered, we have found both translation-based and native solutions for all except CDA and GP. In particular, 
\begin{itemize}
\item TA can be realised by translation into Petri nets~\cite{VVE+06} or SMT solvers~\cite{Grez2015} (Andrea Corradini). The latter is standard in term rewriting (Detlef Plump).
\item WP can be implemented via theorem provers or SMT~\cite{Nolteanalysis, Dragom2019}. There is also work on verification of model transformation using Isabel/HOL~\cite{DBLP:journals/access/BrenasSES18, DBLP:journals/isf/MeghziliCSK19} (Rachid Echahed).
\item SSA can be supported by translation into model checkers. In stochastic model checking we can combine native state space generation with a mapping to Prism or PEPA~\cite{DBLP:conf/ictac/Heckel05,DBLP:journals/sigmetrics/ArijoHTG11}.
\end{itemize}
External tools we considered include theorem provers, SAT and SMT solvers, and model checkers. 

From an engineering perspective of building an analysis tool, we face a range of requirements typical of any software system, which will have to be weighted according to the nature of the analysis problem and the role of the tool, for example as a one-off proof of concept, a long-term platform for research, or an industry-strength analysis tool. Such requirements include external software quality attributes such as functionality, correctness, usability, scalability, etc. as well as internal attributes including maintainability, reusability, and portability. As always, different designs have different characteristics and will address these requirements in different ways.

Building systems by integrating off-the-shelf components is standard practice in software engineering, and it has been argued that it requires a process that is not purely top-down, deriving architectural designs from requirements leading to a specification of the components to be integrated, but an iterative process where top-down needs to be complemented by bottom-up analysis of existing components and subsequent adaptation of the architecture to incorporate such components~\cite{AlbertEvolutionaryProcess2002}. The qualities of the overall system then depend on the architecture, the algorithms and technologies used in the implementation of components developed natively and the qualities of any external components.  

Similarly, factors affecting the characteristics of an analysis tool include (1) the qualities of the external tools and the technologies and languages used for native implementation as well as (2) the characteristics of any translation or analysis algorithm. With respect to (1) there are of course tools at different levels of maturity, more or less stable or future-proof, and with functionalities and performance characteristics that may or may not fit the problem at hand. 

During the discussion at the workshop it was pointed out by Nebras Nassar that mappings to Alloy, but also direct mappings into SAT/SMT solvers, have limited scalability due to to large numbers of variables required for encoding complex model transformations, limiting us to models with roughly 100 to 200 elements. This problem has been investigated for mappings to Alloy, a tool that supports the analysis of software models using SAT/SMT solvers~\cite{semerath2019viatra, varro2018towards}. Generating a logic specification from a given model is a challenge if, due to frequent model updates, the given model has to be translated and checked repeatedly. Such incremental development is common in practice since models are usually created from different perspectives and at different times. While some solvers support incremental analysis, they usually assume monotonic model extensions, not covering scenarios where, for example, models are corrected or repaired. Nassar et al.~\cite{nassar2018ocl2ac} presented the OCL2AC tool, which implements the theory of Habel and Pennemann ~\cite{HP09} and Radke et al.~\cite{radke2018translating} using a native approach. In \cite{nassar2019constructing} the authors also discussed the preservation of invariants and empirically compared the use of application conditions generated with an a-posteriori check of invariants using the facilities provided by the Eclipse Modeling Framework. The tool consists of two components: one for translating OCL constraints to graph constraints, the other for integrating a graph constraint into a transformation rule as a constraint-guaranteeing application condition. 
Sven Schneider confirmed that some solvers are not good at incremental reasoning, especially if this involves not just adding but also removing constraints. 
Detlef Plump and Jens Kosiol argued that theorem provers are more flexible, but often do not provide full automation, and that the encoding of graphs leaves a larger semantic gap, which often shows in the need to handle graph isomorphisms explicitly.

As for (2), many GTS analysis problems are instances of more general problems, e.g., for relational or algebraic structures (including graphs), transitions systems (including graph transition systems), rule-based systems (incl. graph transformation systems), etc. for which general-purpose solutions are available, while others are GTS-specific. The quality of a translation-based solution will depend on the semantic gap between the GTS problem and the problem solved by the (more general) external tool, in particular how closely the translated problem fits into the problem space the external tool is intended for. For example, encoding graphs into logic or algebraic expressions we can end up having many syntactic representations of the same graph, often unrecognisable to the typical user of the tool. 

Hans-J{\"o}rg Kreowski remarked that, even if the tool is fully automatic, as in the case of SAT or SMT solvers, their characteristics may not remain hidden from the user. For example, computer architects use SAT solvers successfully for verifying circuit designs, but when circuits are not verifiable due to scalability issues, they are redesigned to make them easier to verify. That means, models may have to be designed with the algorithm or analysis tool in mind, making an objective comparative evaluation more challenging. 

In conclusion, the problem of assessing different solutions strategies fairly and comprehensively is exacerbated by the many factors influencing the quality of analysis tools, including their architecture, languages, technology and algorithms used, either natively or externally. A possible approach could be to identify a range of benchmarks analysis problems as a basis for a tool competition. The results could give an insight into the design choices of the more successful tools, further informing the debate. 
Such a competition could be organised as part of next GCM editions.  

%
%


\bibliographystyle{eptcs}

\end{document}